\documentclass[twocolumn,showpacs,english,superscriptaddress,prl,preprintnumbers,amsmath,amssymb,floatfix]{revtex4-1}

\usepackage[T1]{fontenc}
\usepackage[latin9]{inputenc}
\usepackage{dcolumn}
\usepackage{bm}
\usepackage{graphicx}
\usepackage{color}
\usepackage{esint}
\usepackage{babel}
\usepackage{amsfonts}
\usepackage{slashed}
\usepackage{enumerate}

\begin{document}
\title{Fermi liquid approach for superconducting Kondo problems}

\author{Alex Zazunov}
\affiliation{Institut f\"ur Theoretische Physik,
Heinrich-Heine-Universit\"at, D-40225  D\"usseldorf, Germany}
\author{Stephan Plugge}
\affiliation{Institut f\"ur Theoretische Physik,
Heinrich-Heine-Universit\"at, D-40225  D\"usseldorf, Germany}
\author{Reinhold Egger}
\affiliation{Institut f\"ur Theoretische Physik,
Heinrich-Heine-Universit\"at, D-40225  D\"usseldorf, Germany}
\date{\today}

\begin{abstract}
We present a Fermi liquid approach to superconducting Kondo problems applicable when the Kondo temperature is large compared to the superconducting gap.
To illustrate the theory, we study the current-phase relation and the Andreev level spectrum for an Anderson impurity between two $s$-wave superconductors.  In the particle-hole symmetric Kondo limit, we find a $4\pi$ periodic Andreev spectrum.  The $4\pi$ periodicity persists under a small voltage bias  which however causes an asymmetric distortion of Andreev levels. The latter distinguishes the present
$4\pi$ effect from the one in topological Majorana junctions.
\end{abstract}

\maketitle

\emph{Introduction.---}The interplay between superconductivity and localized magnetic moments remains of central importance to modern condensed-matter physics.
For instance, spin-fluctuation mediated pairing is encountered in a broad variety of
unconventional superconducting materials \cite{Hewson,Scalapino2012}. Moreover,
Yu-Shiba-Rusinov states induced by a magnetic impurity in a superconductor \cite{Yu1965,Shiba1968,Rusinov1969}
can be responsible for Majorana bound states in magnetic atom chains deposited on  superconducting substrates \cite{Yazdani2014,Franke2015}.
A paradigmatic example for superconducting Kondo problems is given by an Anderson dot in the magnetic regime (where it can realize a Kondo impurity) sandwiched between two conventional $s$-wave BCS superconductors
\cite{Glazman1989,Rozhkov1999,Avishai2003,Vecino2003,Yeyati2003,Siano2004,Choi2004,Sellier2005,
Bergeret2006,Luca2008,Karrasch2008,Meng2009,Andersen2011,Alvaro2011,Luitz2012,Meden2014},
with experimental realizations available in nanoscale devices~\cite{Kasumov1999,VanDam2006,Cleuziou2006,Jorgensen2007,
 Sand-Jespersen2007,Eichler2009,Delagrange2015}.
Numerical calculations  \cite{Siano2004,Choi2004,Karrasch2008,Luitz2012} show
 that the low-temperature physics is governed by the ratio $T_K/\Delta$,
 where $\Delta$ is the superconducting gap and $T_K$ the Kondo temperature (for $\Delta=0$).  While the so-called $\pi$-junction regime with $T_K<\Delta$ is accessible by perturbative renormalization group (RG) methods \cite{Sand-Jespersen2007,Andersen2011},
 the complementary $0$-junction regime with $T_K>\Delta$ has so far withstood analytical progress apart from an exact solution for $T_K/\Delta\to \infty$ \cite{Glazman1989} and different mean-field approximations \cite{Rozhkov1999,Avishai2003,Vecino2003,Yeyati2003,Sellier2005,Bergeret2006, Luca2008,Meng2009}.
 In more general terms, the Kondo effect in a superconductor
represents a long-standing open theoretical problem.

 We here formulate a Fermi liquid theory for the Kondo effect in a superconductor which describes the regime $T_K\gg \Delta$ in a systematic and controlled manner.
For the corresponding normal metal case,  an elegant and asymptotically exact approach has been put forward by Nozi{\`e}res  \cite{Nozieres1974}, cf.~also Refs.~\cite{Hewson,Gogolin2006,Sela2006,Mora2015}. His key insight was that the Kondo singlet formed by the impurity spin and the electron screening cloud can only be polarized, but not broken, near the strong-coupling fixed point.  One then arrives at a Fermi liquid description by expanding the
energy-dependent phase shifts for elastic quasiparticle scattering at low energies and by including residual local quasiparticle interactions \cite{Nozieres1974,Gogolin2006,Sela2006,Mora2015}.
We show below how those ideas can be extended to the superconducting case where, in particular,
 Andreev reflection (AR) processes turn out to be of key importance.  Such processes can be fully captured by a boundary condition accounting both for AR and elastic scattering, cf.~Eq.~\eqref{psm} below.
For $\Delta= 0$, our approach  becomes
equivalent to Nozi{\`e}res' theory. It also reproduces
the $T_K/\Delta\to  \infty$ solution of Ref.~\cite{Glazman1989}.
For a Fermi liquid approach covering the opposite limit $T_K/\Delta\to 0$ in a normal-superconductor junction, see Ref.~\cite{Moca2018}.

\begin{figure}
\centering
\includegraphics[width=0.9\columnwidth]{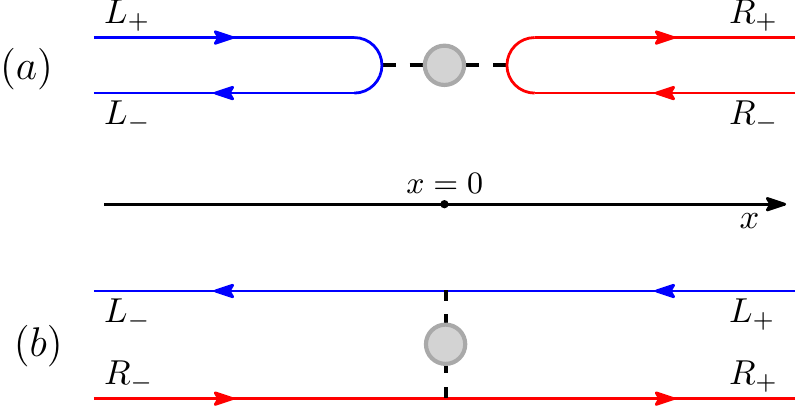}
\caption{Schematic setup. (a) Semi-infinite left/right ($j=L/R$, blue/red) superconducting leads at $x<0/x>0$, respectively, harbor one-dimensional (1D) right ($+$) and left ($-$) movers and are
tunnel coupled (dashed lines) to an Anderson dot (shaded circle) at $x=0$. (b) Unfolded representation with 1D chiral fermions. }
\label{fig1}
\end{figure}

We illustrate our theory for an Anderson impurity between two $s$-wave BCS superconductors, see Fig.~\ref{fig1},  by
studying the Josephson current-phase relation
(CPR), $I(\phi)$, as well as the Andreev level dynamics under a small  bias voltage $V$.
With minor modifications, our theory can be adapted to a
plethora of interesting related problems, e.g., multiple Andreev reflection phenomena
(so far studied only within mean-field schemes \cite{Avishai2003,Yeyati2003}), setups involving
topological superconductors
\cite{Alicea2012,Leijnse2012,Beenakker2013,Mourik2012,Zazunov2016},
or multi-terminal devices \cite{Alvaro2011,Zazunov2017}.
In the particle-hole symmetric Kondo limit of the Anderson model, we predict
a $4\pi$ periodic Andreev level spectrum at low temperature $T\ll \Delta^3/T^2_K$,
with zero-energy level crossings at $\phi=\pi~({\rm mod}~2\pi).$  Such a periodicity is also expected for
 topological Josephson junctions with Majorana states
 \cite{Alicea2012,Leijnse2012,Mourik2012,Beenakker2013,Albrecht2016}
(for experimental signatures, see Refs.~\cite{Bocquillon2017,Laroche2018,Fornieri2018})
and for other setups \cite{Kwon2004,Michelsen2008,Chiu2018}.
We find that under a small bias voltage $V$, the $4\pi$ periodicity persists. However,
 in contrast to all previously studied $4\pi$ periodic setups, the absorption and/or emission spectrum near the zero-energy crossings becomes asymmetric. This fact allows for experimental tests of the underlying mechanism.

\emph{Model.---}We start with an Anderson dot tunnel-coupled to left/right superconducting leads ($j=L/R$), see Fig.~\ref{fig1}(a).
Writing $H = H_d + H_t + H_{\rm leads}$ with
$H_d = \varepsilon_d(n_\uparrow+n_\downarrow)+Un_\uparrow n_\downarrow$, where $n_\sigma= d_\sigma^\dagger d_\sigma^{}$ for dot fermions $d_\sigma$, we have an
interacting ($U>0$) dot level at energy $\varepsilon_d$.
For simplicity taking identical dot-lead tunnel couplings ($t_0$), the point-like tunneling
Hamiltonian is
$H_t = t_0 \sum_\sigma d_\sigma^\dagger b^{}_\sigma(0) + {\rm h.c.}$,
with symmetric
combinations $b_\sigma(x)$ of 1D left/right lead fermion operators, cf.~Eq.~\eqref{ab} below.
Finally, $H_{\rm leads}$ describes $s$-wave BCS superconductor leads \cite{Altland2010}.
Each semi-infinite lead supports right- and left-movers,
$\psi_{j, \sigma}^{(\pm)}(x) \sim e^{\pm i k_F x}$.  In the equivalent unfolded representation
in Fig.~\ref{fig1}(b), we have infinite chiral leads containing only left/right-moving
field operators $\psi_{j,\sigma}(x)$ for lead $j=L/R$, respectively,
$\psi^{(\pm)}_{L, \sigma}(x<0) =  e^{\pm i k_F x} \psi_{L, \sigma}(\mp x)$ and
$\psi^{(\pm)}_{R, \sigma}(x>0) =  e^{\pm i k_F x} \psi_{R, \sigma}(\pm x)$.
To simplify notation, we take the same absolute value $\Delta$ of the superconducting gap
on both sides and put $\hbar = e= v_F = k_B = 1$
(the normal density of states is then just $1/\pi$), resulting in
\begin{eqnarray} \nonumber
H_{\rm leads} &=& \sum_{j=L/R=\pm}  \int_{-\infty}^\infty dx
\Biggl[ \sum_{\sigma = \uparrow, \downarrow} \psi^\dagger_{j, \sigma} \left( \pm i  \partial_x \right) \psi_{j, \sigma} \\  &+&
\Delta \left( e^{\mp i \phi/2} \psi_{j, \downarrow}(x) \psi_{j, \uparrow}(-x) + {\rm h.c.} \right) \Biggr],\label{hleads}
\end{eqnarray}
where $\phi$ is the phase difference.
Next we switch to  the linear combinations
\begin{equation}
\left\{ \begin{array}{c} a_\sigma (x)\\ b_\sigma(x)\end{array}\right\} =
\frac{1}{ \sqrt{2}} \left[ \psi_{L, \sigma}(-x) \mp \psi_{R, \sigma}(x) \right],
\label{ab}
\end{equation}
representing incoming (outgoing) fermion states  for $x<0$ ($x>0$).
The $a$-modes obey open boundary conditions corresponding to
$a_\sigma(0^+) = a_\sigma(0^-)$, which for $t_0=0$ also apply to $b$-modes.

In the magnetic regime, $U \gg \max(\Delta, |t_0|^2)$ and $-U < \varepsilon_d <0$,
the impurity corresponds to a spin-$1/2$ operator ${\bf S}$, with
the particle-hole symmetric Kondo limit at $\varepsilon_d=-U/2$. A
Schrieffer-Wolff transformation yields $H\to H_{\rm leads}+H_{\rm K}$,
where $H_{\rm K}$ contains a potential scattering term (for $\varepsilon_d\ne -U/2$) and an
exchange term with coupling $J>0$ between ${\bf S}$ and the
spin density of $b$-fermions at $x=0$ \cite{Hewson}.
Importantly, $a$-modes always decouple from the impurity and thus can be integrated out
exactly. Using the imaginary-time functional integral approach \cite{Altland2010}, $b$-modes are
then governed by the action $S_b+\int d\tau H_{\rm K}(\tau)$, where
$S_b = - \sum_{k,\omega}\,  \tilde\Psi^\dagger(k, \omega) G^{-1}(k, \omega) \tilde\Psi(k, \omega)$ with
 fermion Matsubara frequencies $\omega$ and
the Nambu spinor
\begin{equation}\label{psib}
\Psi(x, \tau) = \left( \begin{array}{c} b_\uparrow(x, \tau) \\  b^\dagger_\downarrow(-x, \tau) \end{array} \right) \sim
\sum_{k,\omega} e^{i (k x - \omega \tau)}  \tilde\Psi(k, \omega).
\end{equation}
Here and below,  $\tilde\Psi(\omega)$ refers to the frequency representation of
a time-dependent spinor
$\Psi(\tau)$.
After taking into account the pairing-induced bulk coupling between $a$ and $b$ fermions, the free ($t_0=0$) Green's function (GF) appearing in $S_b$
is given by  [cf. Eq.~\eqref{hleads}],
\begin{equation}\label{Gk}
G(k, \omega) = - \frac{i \omega + k \tau_z + \Delta \cos(\phi/2)\tau_x }{ k^2 + \omega^2+\Delta^2} ,
\end{equation}
where Pauli matrices $\tau_{x,z}$ act in Nambu space.

\emph{Weak-coupling regime.---}At high energy scales, the dynamics is restricted to the Hilbert subspace respecting open boundary conditions. Integrating also over the bulk $b_\sigma(x\ne 0)$ modes, we obtain $S_b= - \sum_\omega \tilde\Psi^\dagger(\omega) G^{-1}_0(\omega) \tilde\Psi(\omega)$ with $\Psi(\tau)=\Psi(0, \tau)$ and
\begin{equation}\label{G0}
G_0(\omega) = \int \frac{d k}{2 \pi} \, G(k, \omega) = -
\frac{i \omega + \Delta\cos(\phi/2) \tau_x }{ 2 \sqrt{\omega^2+\Delta^2}}.
\end{equation}
Standard energy-shell integration \cite{Altland2010} then yields the one-loop
RG equations
\begin{equation}\label{RGeqs}
\frac{d J}{ d \ell} = \frac{J^2 }{\pi \sqrt{1 + \delta^2}} ,\quad
\frac{d {\cal Q}}{ d \ell} = - \frac{3}{4 \pi} \frac
{\delta \cos(\phi/2) }{ \sqrt{1 + \delta^2}} J^2,
\end{equation}
where $\delta(\ell) = \Delta/D(\ell)$.
As the effective bandwidth $D(\ell)=e^{-\ell}D$ decreases with increasing RG flow
parameter $\ell$,  a local pairing term,
 $H_{\rm AR}= {\cal Q} b_\downarrow (0) b_\uparrow(0) + {\rm h.c.}$,
 is generated by AR processes.  In fact,
for $\phi\ne \pi$~(mod~$2\pi$), the growing exchange coupling $J(\ell)$ drives ${\cal Q}(\ell)$ toward strong coupling, resulting in Kondo-enhanced AR  \cite{Sand-Jespersen2007,Andersen2011}.
Note that ${\cal Q}(\ell)\sim \cos(\phi/2)$ throughout the flow. However, the RG approach breaks down at energies below
 $T_K\simeq D e^{-\pi/J}$, where one enters the strong-coupling regime.

\emph{Strong-coupling theory.---}In the deep Kondo regime,  the impurity spin is almost
perfectly screened by the leads.  To implement the Fermi liquid approach for the normal case,
it is convenient to employ a scattering state formalism where the leading effects due to the
polarizable Kondo singlet come from energy-dependent phase shifts and residual interaction corrections \cite{Nozieres1974,Gogolin2006,Sela2006,Mora2015}. For the superconducting case, we also need to
 include AR processes. This is achieved below
 by describing both AR and elastic scattering in a unified manner
 through a simple yet general boundary condition.
To that end, by performing a Wick rotation, $i \omega \rightarrow E$,
with energy $E$ relative to the chemical potential $\mu$,
we define $\tilde \Psi_\pm(E)=\tilde\Psi(x=0^\pm,E)$ from the
Nambu spinor \eqref{psib} taken at $x=0^\pm$.
Arbitrary elastic scattering and AR processes are then captured by the boundary
condition
\begin{equation}\label{psm}
\tilde\Psi_+(E) = e^{2 i \hat \eta(E)} \tilde\Psi_-(E), \quad
 \hat \eta(E) = \left( \begin{array}{cc} \eta_\uparrow(E) & \eta_a(E)  \\
-\eta_a^\ast(E) & \eta_\downarrow(-E) \end{array} \right),
\end{equation}
 where the Nambu matrix $\hat \eta(E)$ has the most general form allowed by
 Hermiticity of the self-energy $\hat\Sigma(E)$ in
 Eq.~\eqref{selfenergy} below.
 While the real functions $\eta_{\uparrow,\downarrow}(E)$ are
 energy-dependent phase shifts precisely as in the normal case,
the complex-valued function $\eta_a(E)$ describes AR.

Next, Eq.~\eqref{psm} is linked to the retarded response of bulk modes,
$\tilde\Psi_\pm( E) = \sum_k e^{i k 0^\pm} G^R(k, E) \tilde\Psi(E)$, to an
effective boundary field, $\tilde\Psi(E)$, living at $x=0$.
Using the retarded GFs $G^R(k,E)$ and $G^R_0(E)$ obtained by  Wick rotation
from Eqs.~\eqref{Gk} and \eqref{G0}, respectively, we find
$\tilde\Psi_\pm(E)=\left( G_0^R(E) \mp \frac{i}{ 2} \tau_z\right) \tilde\Psi(E).$
Here the $\tau_z$ term originates from the respective $\tau_z$ term in Eq.~\eqref{Gk}.
 One can thereby write
Eq.~\eqref{psm} as equation of motion for the boundary spinor,
\begin{equation}\label{selfenergy}
\left[ G_0^R(E)+ \hat\Sigma(E) \right] \tilde\Psi(E) = 0, \quad
\hat\Sigma(E) = \frac{1}{2}\cot[\hat \eta(E)] \tau_z.
\end{equation}
Finally passing back to imaginary time and rescaling $\Psi(\tau)=
\frac{1}{\sqrt{2}} \left(  b_\uparrow(\tau),  b^\dagger_\downarrow(\tau)\right)^T$,
the strong-coupling action is given by [cf.~Eqs.~\eqref{G0} and \eqref{selfenergy}]
\begin{eqnarray}\label{scaction}
S_{\rm sc}[\Psi]&=&- \sum_\omega \tilde\Psi^\dagger(\omega) {\cal G}^{-1}(\omega)
 \tilde\Psi(\omega)+S_I, \\
 \nonumber
{\cal G}^{-1}(\omega) &=& {\cal G}_0^{-1}(\omega) -
\cot[\hat \eta(i \omega)] \tau_z,  \quad {\cal G}_0^{-1}(\omega)=-2G_0(\omega),
 \end{eqnarray}
while $S_{I}$ describes residual interaction corrections addressed below.
We emphasize that our self-energy formulation of AR and elastic scattering processes in
Eq.~\eqref{scaction} is  completely general.

In order to arrive at a low-energy Fermi liquid theory, we now expand $\hat \eta(E)$ in powers of $|E|/T_K\ll 1$ and $\Delta/T_K\ll 1$.
Using the spin symmetry of the problem and noting that conventional even-frequency pairing generated from Eq.~\eqref{hleads} implies $\eta_a(-E) = \eta_a(E)$, we find
\begin{eqnarray}\label{expan}
 \eta_\uparrow(E) &=& \eta_\downarrow(E)=\eta_F + \alpha_1 E + \alpha_2 E^2 + \cdots,\\
\nonumber
\eta_a(E) &=& \Delta \left( \beta_1 + \beta_3 E^2 + \cdots \right),
\end{eqnarray}
where $\eta_F$ is the quasiparticle phase shift at the Fermi energy for $\Delta = 0$.
The Fermi liquid parameters $\alpha_n$ and $\beta_n$ scale as $1/T_K^n$, where
the $\alpha_n$ determine the elastic scattering
phase shifts  \cite{Nozieres1974,Mora2015} and
the complex-valued $\beta_n$ depend on the phase difference $\phi$ (see below).
Keeping all terms up to order $1/T_K^2$, and using the renormalized  parameters $\tilde \alpha_{n} =   \alpha_n/\sin^2 \eta_F$ and
$\tilde \beta_n =\beta_n/  \sin^2 \eta_F$,
 we arrive at
\begin{eqnarray}\nonumber
{\cal G}^{-1}(\omega) &=& {\cal G}_0^{-1}(\omega)-
\left( \begin{array}{cc} \lambda(\omega)-i\tilde\alpha_1\omega & \tilde\beta_1 \Delta
 \\ \tilde\beta_1^\ast \Delta& -\lambda(\omega)-
i\tilde\alpha_1 \omega  \end{array} \right),    \\   \label{ss0}
 \lambda(\omega)& =&  \cot \eta_F\left(
1-\frac{\alpha_1^2\omega^2+|\beta_1|^2\Delta^2}{\sin^2\eta_F}\right)
+\tilde\alpha_2  \omega^2.
\end{eqnarray}
Further simplifications arise in the Kondo limit, where particle-hole symmetry (which is not broken by pairing terms) imposes the condition $\tau_xe^{2i\hat \eta(E)}\tau_x=e^{-2i\hat\eta(E)}$ \cite{SM},
resulting in $\eta_F=\pi/2$, $\alpha_2=0$, and $\beta_1=\beta^*_1$.  In the Kondo limit, we thus
have $\lambda(\omega)=0$ in Eq.~\eqref{ss0}.

\emph{Residual interaction processes.---}We now turn to $S_I$ in Eq.~\eqref{scaction}.
Keeping all terms up to order $1/T_K^2$, this action contribution has the general form
\begin{equation}\label{Sint}
S_I = \frac12\sum_{\sigma=\uparrow,\downarrow} \int d \tau \
 b^\dagger_{-\sigma} b_{-\sigma}  b^\dagger_\sigma
\left( \tilde u_1 - \tilde u_2 \partial_\tau \right) b_\sigma,
\end{equation}
with expansion parameters $\tilde u_n\sim 1/T_K^n$ (where $\tilde u_1\ge 0$).
Defining normal ordering and averages $\langle\cdots\rangle_0$ with
respect to  the BCS ground state for
${\cal G}_0(\omega)$, cf.~Eq.~\eqref{scaction}, it is convenient to
express Eq.~\eqref{Sint} by
virtue of Wick's theorem as $S_I=\langle S_I\rangle_0 +\tilde S_I+ S_I^H,$
where $\tilde S_I$ is the normal-ordered form of Eq.~\eqref{Sint} and
 $S_I^H$ represents Hartree terms
which can be accounted for via the $\hat\eta (E)$-expansion
in Eq.~\eqref{expan}.  Up to order $1/T_K^2$, with $u_n=\tilde u_n \sin^2\eta_F$, we find
\begin{eqnarray}
\eta_\sigma(E) &=& \eta_F + \alpha_1 E +
\alpha_2 E^2 - \left( u_1 + u_2 E \right)\delta N_{-\sigma},   \nonumber\\
\eta_a(E) &=& \beta_1  \Delta + u_1 \delta Q,\label{etaexp}
\end{eqnarray}
where $\delta N_\sigma$ and $\delta Q$ are self-consistent Hartree parameters for
local density and pairing fluctuations, respectively.
Again invoking spin symmetry,  $\delta N_\uparrow = \delta N_\downarrow$,  Eq.~\eqref{etaexp} implies that Hartree terms can indeed
 be included by renormalizing
$\alpha_{n}$ and $\beta_n$. We assume henceforth that this renormalization has already
been carried out.  Moreover, since the Kondo singularity is tied to the Fermi level,
the phase shifts $\eta_\sigma(E)$ must be independent of the chemical potential $\mu$ \cite{Nozieres1974,Mora2015}.
This fact implies that one can derive
relations between Fermi liquid parameters
without having to specify $\delta N_\sigma$ or $\delta Q$
\cite{Nozieres1974,Mora2015}.
In particular, in the Kondo limit, $\partial_\mu\eta_F=0$ and $\alpha_2=u_2=0$ imply the well-known identity $u_1=\pi\alpha_1$ \cite{Nozieres1974} and $\partial_\mu\alpha_1=0$.

\emph{Current-phase relation.---}The CPR follows as phase derivative of the free energy,
\begin{equation} \label{cprgen}
I(\phi) = 2\partial_\phi F= I_A(\phi)+ I_{\rm int}^{(1)}(\phi)+I_{\rm int}^{(2)}(\phi),
\end{equation}
where $I_A(\phi)=-2 T\sum_\omega \partial_\phi \ln\det {\cal G}^{-1}(\omega)$ is the
Andreev bound state (ABS) contribution, see Eq.~\eqref{ss0}. In particular, the ABS
spectrum follows by solving det$[{\cal G}^{-1}(-iE)]=0$ for subgap energies, $|E|<\Delta$.
Keeping terms up to order $1/T_K$, where $\lambda(-iE)=\lambda=\cot\eta_F$ [cf.~Eq.~\eqref{ss0}], this condition reads
\begin{equation}\label{ABS}
 \frac {E^2}{\Delta^2}= \frac{\left|\cos(\phi/2)-\tilde\beta_1
\sqrt{\Delta^2-E^2}\right|^2+\lambda^2}
{\left(1+\tilde\alpha_1\sqrt{\Delta^2-E^2}\right)^2+\lambda^2}.
\end{equation}
In the Kondo limit (with $\lambda=0$), Eq.~\eqref{ABS} holds up to
order $1/T_K^2$.

The leading interaction contribution to the CPR, see Eq.~\eqref{cprgen}, follows from $\langle S_I\rangle_0$  \cite{SM},
\begin{equation}\label{I1}
I^{(1)}_{\rm int} (\phi)=\delta I_c \sin\phi,\quad
\delta I_c\simeq - \frac{\tilde u_1\Delta^2}{4\pi^2}\ln^2 \left(T_K / \Delta \right).
\end{equation}
As expected in the presence of repulsive quasiparticle interactions, we obtain a decrease of the critical current, $\delta I_c<0$, where $|\delta I_c| \sim \frac{\Delta^2}{T_K}\ln^2(T_K/\Delta)$ contains a logarithmic enhancement factor.
Finally, $I_{\rm int}^{(2)}$ describes higher-order interaction corrections to the CPR
due to $\tilde S_I$.
To order $1/T_K^2$, we obtain \cite{SM}
\begin{equation} \label{I2}
I^{(2)}_{\rm int}(\phi) \approx \tilde u_1^2 \Delta^3
\left( \sin \phi+\frac{1}{ 2}\sin(2 \phi) \right),
\end{equation}
where the $\sin (2 \phi)$ term describes coherent tunneling processes involving two Cooper pairs.

\begin{figure}[t]
\centering
\includegraphics[width=0.95\columnwidth]{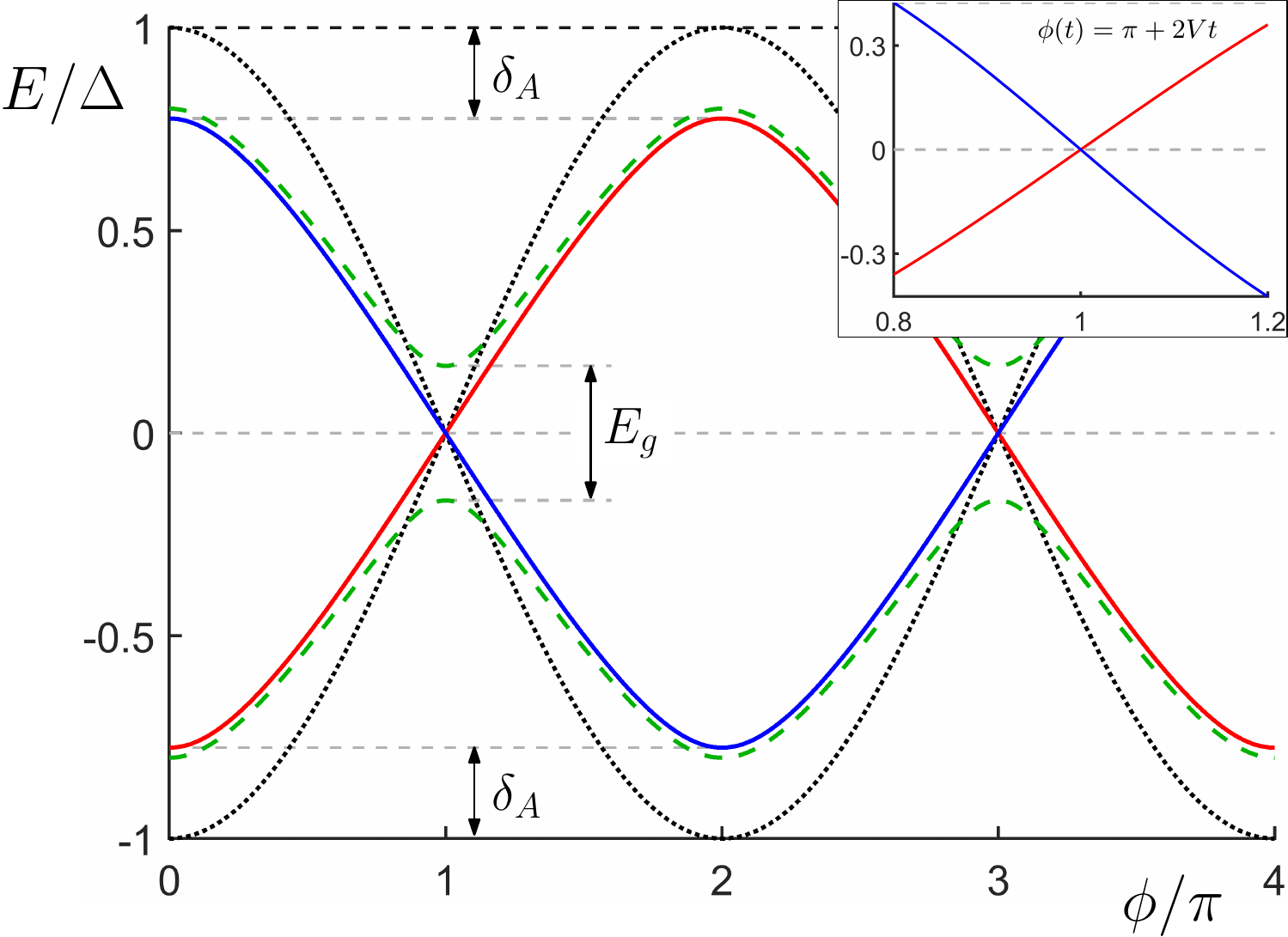}
\caption{ABS spectrum vs phase $\phi$.  Main panel: Black dotted curves show the particle-hole symmetric limit with $T_K/\Delta\to \infty$ \cite{Glazman1989}.  Blue and red solid curves
depict $4\pi$ periodic Andreev levels for $T_K/\Delta=5, \lambda=0$, and $\alpha_1=\gamma=1/T_K$.  Green dashed curves illustrate the gap $E_g$ formed away from particle-hole symmetry ($\lambda=0.2$), leading to $2\pi$ periodicity.
Inset:  Asymmetry of $4\pi$-periodic adiabatic Andreev levels near the crossing at $\phi=\pi$ with voltage  $V=0.33\Delta$ and ${\rm Im}(\gamma)=3/T_K$.  }
\label{fig2}
\end{figure}

Let us then turn to the dominant ABS contribution, see Eq.~\eqref{ABS}, where the $\phi$-dependence of the AR coupling $\tilde\beta_1$ follows from Eq.~\eqref{RGeqs},
$\tilde\beta_1(\phi)=\gamma \cos(\phi/2),$
with constant $\gamma\sim 1/T_K$.
    (i)  For $T_K/\Delta\to \infty$, all Fermi liquid parameters and thus also the interaction corrections \eqref{I1} and \eqref{I2} can be dropped. Solutions to Eq.~\eqref{ABS} are then given by
$E=\pm \Delta\sqrt{1-{\cal T}\sin^2(\phi/2)}$ with
the junction transparency ${\cal T} = \sin^2\eta_F=1/(1+\lambda^2)$.
We thus readily recover the results of Ref.~\cite{Glazman1989}.
(ii) Including $1/T_K$ corrections, see Fig.~\ref{fig2}, Eq.~\eqref{ABS} predicts a $4\pi$ periodic ABS spectrum in the Kondo limit ($\lambda=0$), with zero-energy ABS crossings at $\phi=\pi~({\rm mod}~2\pi$).
For $\lambda\ne 0$, we instead have avoided crossings with gap $E_g\simeq 2\sqrt{1-{\cal T}}\Delta$,
and thus obtain a conventional $2\pi$ periodic spectrum.
(iii) Fermi liquid corrections imply a detachment of ABSs from quasiparticle continuum states
at $\phi=0~({\rm mod}~2\pi)$. The detachment gap, $\delta_A=\Delta-E_A(0)$, follows from Eq.~\eqref{ABS} as
\begin{equation}\label{detach}
\delta_A = 2\sin^4(\eta_F)\Delta^3
\left[ \tilde\alpha_1+{\rm Re}(\gamma)\right]^2\sim
\frac{ \Delta^3}{T_K^2}.
\end{equation}
While ABS detachment already arises from elastic scattering \cite{Yeyati2003}, AR and
Hartree corrections can strongly renormalize $\delta_A$.
Since the Kondo resonance floats with the Fermi level and the
ABS spectrum is detached from the continuum, the $4\pi$ periodic CPR in the Kondo limit
should be observable for $T\ll \delta_A$.

\emph{ABS spectrum for small voltage $V$.---}What will happen to the $4\pi$ periodic Andreev spectrum in the Kondo limit when a small bias voltage $V$ is applied?  For $V\ll \delta_A\ll \Delta$, adiabatic Andreev levels still represent good dynamical variables. Since the ABSs are removed from continuum states by a spectral gap,  the retarded and advanced sectors of the Keldysh action
decouple \cite{Altland2010,SM}.  To investigate whether the $4\pi$ periodicity survives in the nonequilibrium case, we consider the phase dynamics, $\phi(t)=\pi+2Vt$, at times where $\phi(t)\approx \pi~({\rm mod}~2\pi)$, corresponding to zero-energy crossings.   The retarded sector can equivalently be described \cite{SM} by the real-time action
\begin{equation}\label{retaction}
S=  \int dt  \ \Phi^\dagger(t) \left[ i\partial_t - E_A(t) \left(\sigma_z-\xi(t) \right) \right] \Phi(t),
\end{equation}
where $\Phi=(c_+,c_-)^T$ contains the amplitudes for
upper/lower ($\nu=+/-$) Andreev branches, the Pauli matrix $\sigma_z$ acts in Andreev level space, and
\begin{equation}\label{ea1}
E_A(t)\simeq \frac{1-{\rm Re}(\gamma)\Delta}{1+\alpha_1\Delta}\ \Delta\sin(Vt) .
\end{equation}
In the near-adiabatic regime, the ABS degeneracy at each crossing is not lifted by the voltage,
and the Andreev spectrum remains $4\pi$ periodic.
We find $\xi(t)=\frac12 {\rm Im}(\gamma) V\cos(Vt)$ in Eq.~\eqref{retaction}, where ${\rm Im}(\gamma)\ne 0$ requires that particle-hole symmetry has been broken, e.g., by the voltage.
The only effect of $\xi(t)$ then consists of an asymmetric distortion of the $\nu=+/-$ Andreev levels, cf.~Eqs.~\eqref{retaction}, \eqref{ea1} and inset of Fig.~\ref{fig2},
\begin{equation}\label{asymmetry}
\nu E_A \to \nu (1-\nu \xi) E_A.
\end{equation}
Since $\xi\to -\xi$ at each subsequent crossing ($\phi\to \phi+2\pi$),
Eq.~\eqref{asymmetry} implies an asymmetric absorption and/or emission spectrum near the ABS crossings.  Importantly, this feature allows one to experimentally distinguish the predicted $4\pi$ Josephson effect  from its topological counterpart in Majorana junctions \cite{Alicea2012,Leijnse2012,Beenakker2013} as well as from other proposed realizations \cite{Kwon2004,Michelsen2008,Chiu2018}.
The real and imaginary parts of $\gamma$ can be measured via the detachment gap $\delta_A$ [Eq.~\eqref{detach}] in the equilibrium Andreev spectrum and via the low-voltage asymmetry $\xi$, see Eq.~\eqref{asymmetry}, respectively.

\emph{Conclusions.---}In this work, we have presented a Fermi liquid approach to the Kondo problem in a conventional $s$-wave BCS superconductor with $T_K\gg \Delta$.
 While we have illustrated the theory for
an Anderson dot between two superconducting leads in the (near) equilibrium regime,
 the Fermi liquid description also allows to tackle many other setups featuring an interplay of Kondo physics with superconductivity.

\begin{acknowledgments}
We thank K. Flensberg, A. Levy Yeyati, and F. von Oppen for discussions and
acknowledge funding by the Deutsche Forschungsgemeinschaft (Bonn),
 Grant No.~EG 96/11-1.
\end{acknowledgments}


\newpage
\begin{appendix}
\setcounter{equation}{0}
\renewcommand{\theequation}{S\arabic{equation}}

\begin{center}
\textbf{\large Supplemental Material to ``Fermi liquid approach for superconducting Kondo problems''}
\end{center}

\vspace{.3cm}
We here provide details about particle-hole symmetry constraints as
well as short derivations of Eqs.~(16), (17) and (19) quoted in the
main text.

\subsection{Particle-hole symmetric Kondo limit}

First we address the derivation of the relation
\begin{equation}\label{phs1}
\tau_xe^{2i\hat \eta(E)}\tau_x=e^{-2i\hat\eta(E)},
\end{equation}
which holds in the particle-hole (PH) symmetric Kondo limit of the Anderson model
with $\hat\eta(E)$ in Eq.~(7) of the main text.
The PH transformation ${\cal P}$ amounts to exchanging
$ \tilde b_\sigma(x,E)\leftrightarrow
\tilde b^\dagger_{-\sigma}(x,-E)$
 such that
${\cal P}\tilde\Psi_\pm(E)=\tau_x\tilde\Psi_{\mp}(E).$
By virtue of  the relation
\[
\tau_x \left[ G_0^R(E)\mp\frac{i}{2}\tau_z\right]\tau_x= G_0^R(E)\pm \frac{i}{2}\tau_z,
\]
we find that the bulk action is  ${\cal P}$-invariant. Concerning the boundary condition [cf.~Eq.~(7) in the main text], ${\cal P}$-invariance implies
 the condition
${\cal P}\tilde\Psi_+(E)=e^{2i\hat \eta(E)}{\cal P}\tilde \Psi_-(E).$
Hence we obtain Eq.~\eqref{phs1}.

\subsection{Interaction corrections}

Here we give additional details on the derivation of Eqs.~(16) and (17) in the main text,
where the leading interaction contributions $I^{(1)}_{\rm int}(\phi)$ and $I^{(2)}_{\rm int}(\phi)$, respectively, have been specified.
 First, the anomalous Hartree contribution to $\langle S_I\rangle_0$ follows from
\begin{eqnarray}\nonumber
\langle b_\downarrow(\tau)b_\uparrow(0)\rangle_0 &\simeq& \frac{\Delta\cos(\phi/2)}{2\pi}
\int_0^{T_K} \frac{d\omega}{\sqrt{\omega^2+\Delta^2}} \\ &\simeq &
\frac{\Delta\cos(\phi/2)}{2\pi} \ln(T_K/  \Delta).
\end{eqnarray}
The corresponding free energy contribution from $\langle S_I\rangle_0$ is then given by
\begin{equation}
F^{(1)}_{\rm int}(\phi)=\tilde u_1 (\Delta/2\pi)^2\ln^2(T_K/\Delta) \cos^2(\phi/2)
\end{equation}
and yields Eq.~(16) in the main text.
Second, higher-order corrections follow by cumulant expansion in $\tilde S_I$,
\begin{equation}\label{f11}
e^{-F^{(2)}_{\rm int}/T} = \left\langle {\cal T}_\tau e^{-\tilde S_I} \right\rangle = \exp\left(\frac12\langle \tilde S_I^2\rangle_c + \cdots\right),
\end{equation}
where $\langle\cdots\rangle_c$ indicates that only connected diagrams are included and ${\cal T}_\tau$ is the imaginary-time ordering operator. To order $1/T_K^2$, we find from Eq.~\eqref{f11} the contribution
\begin{equation}
F^{(2)}_{\rm int} = -\frac{T\tilde u_1^2}{2}\int_0^{1/T} d\tau_1 d\tau_2 \langle B(\tau_1)
B( \tau_2) \rangle_c
\end{equation}
with $B(\tau)= b^\dagger_\uparrow(\tau) b^{}_\uparrow(\tau) b^\dagger_\downarrow(\tau) b^{}_\downarrow(\tau)$.   Here $\phi$-dependent contributions mainly originate from the diagram with four anomalous contractions while the diagram with four normal contractions depends only weakly on $\phi$ and can be neglected.  As a result, taking $T\to 0,$ we obtain
\begin{eqnarray}
F^{(2)}_{\rm int}&= & -A\cos^4(\phi/2),\\ \nonumber
A&\simeq& \tilde u_1^2   \Delta^3 \int_0^\infty
d\xi \left( \int_0^{T_K/\Delta} dx \frac{\cos(x\xi)}{\pi\sqrt{1+x^2}}\right)^4 ,
\end{eqnarray}
which yields Eq.~(17) quoted in the main text.

\subsection{Adiabatic Andreev levels at small voltage}

Consider the case of low bias voltage, $V \ll \delta_A \ll \Delta$,
which implies a slowly varying phase difference, $\dot \phi(t) = 2 V$.
The ABS occupation dynamics then stays almost all the time away from the gap edges
such that the retarded and advanced sectors of the full Keldysh action are decoupled during the time evolution.
The subgap dynamics is thus already described by an effective action for the retarded sector,
\begin{equation}\label{lowV:S}
S = \int dt dt' \Psi^\dagger_q(t) {\cal L}(t, t') \Psi_c(t'),
\end{equation}
with
\begin{equation}\label{lagra}
    {\cal L}(t,t') = \frac12 \left( G_+ + G_- \right) - \hat\Sigma.
\end{equation}
Here we have defined
\begin{equation}\label{gf1}
G_\pm(t,t') = e^{\pm i \tau_z \phi(t)/4} G_0^R(t-t') e^{\mp i \tau_z \phi(t')/4}
\end{equation}
and the Fourier transform of $G_0^R(t)$ is given by
\begin{equation}
G_0^R(E) = \frac{E + \Delta \tau_x }{\zeta_E},\quad \zeta_E=\sqrt{\Delta^2-(E+i0^+)^2}.
\end{equation}
The Nambu spinors $\Psi_c$ and $\Psi^\dagger_q$ are the 'classical' and 'quantum' components of
the boundary-field Keldysh spinor, respectively. For ease of notation, we drop the indices $(c,q)$ in what follows.

First, in the \emph{adiabatic} approximation, one neglects $\dot \phi^2 \sim V^2$, $\ddot \phi$, and all higher-order time derivatives.
The GFs in Eq.~\eqref{gf1} then take the form
\begin{eqnarray}
\label{lowV:Gpm}
&& G_\pm(t,t') \simeq \\
&& \nonumber \frac{1}{\zeta(t)} \left( \pm \frac{\dot \phi(t)}{4} \tau_z + i \partial_t +
\Delta e^{\pm i \tau_z \phi(t)/2} \tau_x \right) \delta(t-t').
\end{eqnarray}
In addition, one puts $\zeta(t) = \sqrt{\Delta^2 - E_A^2(t)}$
with the instantaneous ABS energy $E_A(t) \equiv E_A(\phi(t))$, where $E_A(\phi)$ solves the equilibrium condition
 in Eq.~(15) of the main text. After rescaling
$\Psi(t) \to \sqrt{\frac{\zeta(t)}{ 1 + \tilde \alpha_1 \zeta(t)}} \Psi(t),$
 the effective action, $S = \int dt \Psi^\dagger(t) {\cal L}(t) \Psi(t)$, has the time-local Lagrangian
\begin{eqnarray}\nonumber
{\cal L}(t) &=& i \partial_t + \frac{\Delta}{1 + \tilde \alpha_1 \zeta(t)}
\left[ \cos\left(\frac{\phi(t)}{2} \right)\tau_x - \hat \beta_1(t) \zeta(t) \right] \\ && -
\frac{\lambda \zeta(t)}{1 + \tilde \alpha_1 \zeta(t)} \tau_z, \label{lowV:Sad}
\end{eqnarray}
where $\hat \beta_1=\left( \begin{array}{cc}0&\tilde\beta_1\\ \tilde\beta_1^\ast & 0\end{array}\right)$ and the time dependence of $\tilde \beta_1$ follows from the time dependence of the phase.

A systematic way to compute \emph{nonadiabatic corrections} is to expand Eq.~\eqref{lowV:S} in powers of $\partial_t$,
\begin{eqnarray}
&& S = \int dt d\tau \Psi^\dagger(t + \tau/2) {\cal L}(t; \tau) \Psi(t - \tau/2), \\
\nonumber
&& {\cal L}(t+\tau/2, t-\tau/2) \equiv {\cal L}(t; \tau) =
\int\frac{d E}{ 2\pi} \, e^{-i E \tau} {\cal L}(t; E),
\end{eqnarray}
where $t$ is the 'center-of-mass' (and $\tau$ the relative) time.  The Lagrangian, see Eq.~\eqref{lagra}, in this mixed representation,
${\cal L}(t; E)$, involves the GF matrices [see Eq.~\eqref{gf1}]
\begin{equation}
G_{s=\pm}(t; E) = \frac{E + \tau_z sV/2 }{ \zeta_{E + \tau_z sV/2}} + \frac{\Delta}{ \zeta_E}
\left( \begin{array}{cc} 0 & e^{i s Vt} \\ e^{-i s Vt} & 0 \end{array} \right),
\end{equation}
and the self-energy part is given by
\begin{equation}
\hat\Sigma(t; E) \approx \lambda \tau_z - \tilde \alpha_1 E +
\Delta \left( \begin{array}{cc} 0 & \gamma \\ \gamma^\ast & 0 \end{array}\right)
\cos(Vt) .
\end{equation}
For a low-energy description, we neglect continuum states by projecting $\Psi(E) \rightarrow \Theta(\Delta - |E|) \Psi(E)$,
which is justified for $|E \pm V/2|<\Delta$. We note that ${\cal L}(t; E)$ then also stays Hermitian.
Since ${\cal L}(t; E)$ only slowly depends on $t$, to leading nontrivial order, the action is given  by
\begin{eqnarray}\nonumber
S &=& \frac12 \int dt \Psi^\dagger(t) \left[  {\cal L}(t; E) +
\frac{i}{2}\partial_t \partial_E {\cal L}(t; E) \right]_{E = i \partial_t} \Psi(t)\\ && + {\rm h.c.},\label{lowV:Sexp}
\end{eqnarray}
where we neglect terms $\sim\partial_t^n {\cal L}(t; E) \propto V^n$ with $n \geq 2$.

We next introduce Nambu spinor eigenstates, $\chi_{\nu = \pm}(t)$, for instantaneous Andreev levels,
\begin{equation}
{\cal L}(t; E = \nu E_A(t)) \chi_\nu(t) = 0,\quad
\det\left[ {\cal L}(t; E = \nu E_A(t))\right] = 0,
\end{equation}
with $\chi^\dagger_\nu(t) \cdot \chi_{\nu'}(t) = \delta_{\nu \nu'}$. Expanding $\Psi(t)$ in this adiabatic Andreev basis,
\begin{equation}\label{lowV:Psiad}
\Psi(t) = \sum_{\nu = \pm} c_\nu(t) \chi_\nu(t) e^{-i \nu \int^t ds E_A(s)},
\end{equation}
and substituting Eq.~\eqref{lowV:Psiad} into Eq.~\eqref{lowV:Sexp},
the effective action is written in terms of the amplitudes $c_\nu(t)$,
\begin{eqnarray}\label{lowV:Snonad}
S &= &\frac{i }{ 2} \int dt \sum_{\nu, \nu' = \pm}  e^{i (\nu' - \nu) \int^t ds E_A(s)}
c_{\nu'}^\dagger(t) \chi^\dagger_{\nu'}(t)  \\ \nonumber &\times&\left[
\partial_E {\cal L}(t; \nu E_A) \partial_t + \frac12\partial_t \partial_E {\cal L}(t; \nu E_A)
\right] c_\nu(t) \chi_\nu(t)  + {\rm h.c.}
\end{eqnarray}

We now focus on the vicinity of the Kondo limit, where $\lambda = 0$. However, $\beta_1$ can now be complex-valued since we allow for
particle-hole symmetry breaking. In the mixed representation, cf.~Eq.~\eqref{lowV:Sad}, we find
\begin{equation}
{\cal L}(t; E) = E + w_E \Delta \cos[\phi(t)/2] \left( \begin{array}{cc} 0 & e^{- i \theta_E}  \\
e^{i \theta_E} & 0 \end{array} \right),
\end{equation}
where $w_E = \frac{\left| 1 - \gamma \zeta_E \right|}{1 + \alpha_1 \zeta_E}$
and $\theta_E = \tan^{-1} \frac{{\rm Im}(\gamma) \zeta_E}{1 - {\rm Re}(\gamma) \zeta_E}$.  We here
assume $\phi(t) \equiv \pi + 2 V t \approx \pi \,{\rm mod}(2 \pi)$,
i.e., we are near an ABS crossing. The adiabatic ABS energies follow as
$E(t) = \pm w_E \Delta \sin(Vt)$, cf.~ Eq.~(20) of the main text.
The corresponding eigenstates are $\chi_{\nu = \pm}(t) = (1, \nu e^{i \theta(t)})^T/\sqrt{2}$
with $\theta(t) = \theta_{E_A(t)}$.
Substituting these expressions into Eq.~\eqref{lowV:Snonad}, computing the matrix elements between
the states $\chi_\nu(t)$, and finally passing to the Heisenberg picture,
$c_\nu(t) \to e^{i \nu \int^t ds E_A(s)}  c_\nu(t)$,
we arrive at the action specified in Eq.~(19) of the main text.
\end{appendix}

\end{document}